\documentclass{class}

\usepackage{graphicx}
\usepackage{amsmath}
\usepackage{amsfonts}
\usepackage{amssymb}

\title{On the Structure of Tree:- A conceptual article}

\author{MUNISHWER CHANDER CHADDA$^{1}$ AND MAYANK CHADHA$^{2}$}

\heading{On the structure of tree: Munishwer Chander Chadda and Mayank Chadha}

\address{$^{1}$ Kailash Engineering Company\\
22, Deshmesh Market, Port Blair 744101\\
Andaman and Nicobar Island, India\\
chadhakailash25@gmail.com
\and
$^{2}$ University of California, San Diego\\
9500 Gilman Drive, La Jolla CA 92092-0085\\
machadha@ucsd.edu
}

\keywords{Tree, Optimization, and Geometry}

\abstract{This article is a conceptual exposition on the structure of the tree. It demonstrates an evolutionary design that the tree possesses in the perspective of a structural engineer.}

\begin{document}

\thispagestyle{empty}

\section{INTRODUCTION AND MOTIVATION}

Mankind has always been inspired by nature. Every single unit of nature, including the universe, is a structure. Birds, human beings, trees, the solar system and volcanoes are some examples of dynamic structures, whereas certain structures like hills, mountains and rocks can be grouped under static structures. The evolutionary character of nature, in our opinion, is directly related to continuous design optimization governed by immediate environmental influence. Therefore, evolution in nature can be looked at as a design optimization problem. For instance, in the study on the aerodynamics of gliding flight in a Falcon,  Tucker et al. \cite{tucker:1970} showed that the falcon adjusted its wing span in flight to achieve nearly the maximum possible lift to drag ratio over its range of gliding speeds. The optimum design of the elements in nature is sufficient motivation to analyze a tree as a structure subjected to environmental loads.

This paper focuses on the conceptual analysis of a tree structure as an epitome of both material and geometric optimization. This, in turn, generates enough motivation to adapt the design ideas from nature.

\section{STRUCTURE OF THE TREE}
A tree can be simplified into the form of a cantilever beam fixed at the bottom subjected primarily to gravity load (its own self-weight), wind loads and earthquake loads. A tree like any slender structure, can be considered as a configuration in three-dimensional space that can be described by locus of geometric centroid (called the midcurve) and the family of cross-section attached to this midcurve. The tree can be structurally appreciated by observing the optimization of this configuration, the material properties and the resistance to the given loads. We discuss these parameters in detail in the following paragraphs.
\subsection{On geometry and load}
\begin{figure}[ht]\centering 
\includegraphics[width=\textwidth]{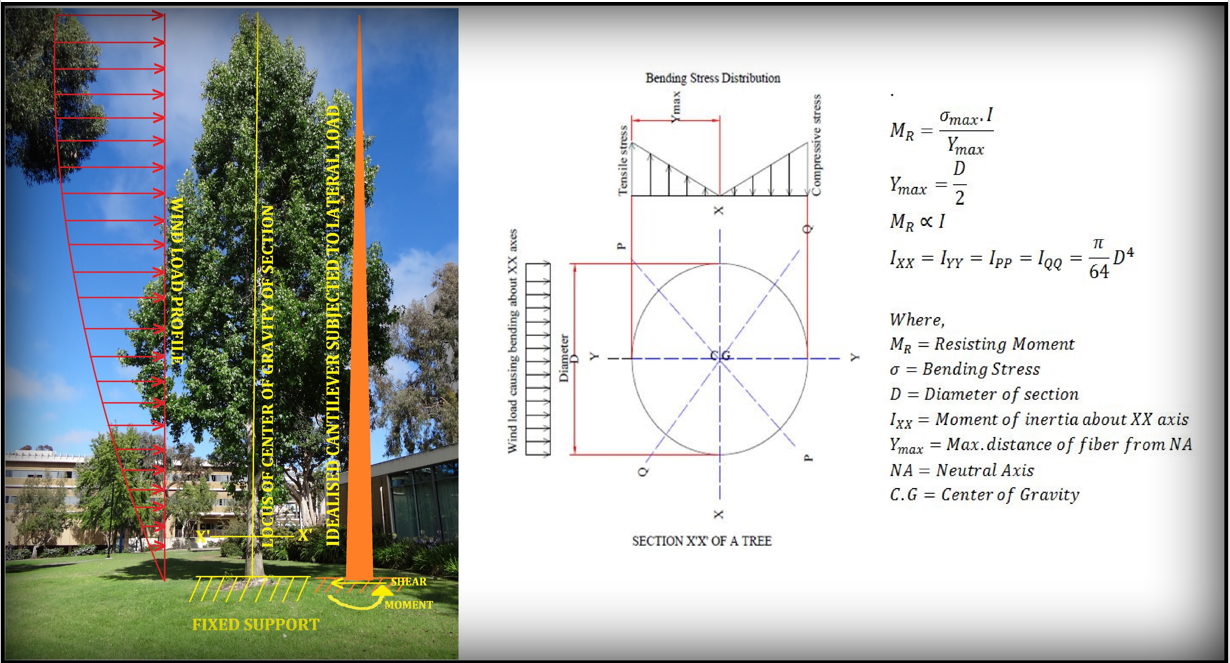}
\caption{Tree idealized as cantilever beam subjected to lateral loads and axial compression due to self weight.}
\label{fig:Figure1}
\end{figure}

Consider a typical structure of tree. In an ideal situation, it can be observed that the cross-section of the tree is circular. A circle is the only cross-section in geometry that has an equal moment of inertia about all axes on the horizontal plane. This in turn, implies that the moment of resistance about all the axes in the horizontal plane is the same. This is of major advantage to the tree. The circular (or very closely circular) cross-section serves for the requirement to resist laterally acting wind load in any direction of wind flow or any lateral force like earthquakes. Most of the mass of the tree is concentrated in the trunk. Therefore, the midcurve of the tree is nearly a straight line, maintaining symmetry, thereby avoiding eccentricity. Figure \ref{fig:Figure1} illustrates the idea above. 

The tree trunk has a larger diameter at the bottom, and the cross-sectional area reduces gradually as we move towards the tip of the tree. There is a peculiar advantage of this configuration. Firstly, it lowers the center of mass of the tree closer to its bottom. This helps prevent overturning against lateral loads. Secondly, since the primary load that the tree must resist is lateral load (wind and earthquake load), the section of the tree that will resist the maximum moment is at the bottom, and the larger cross-section at the bottom guarantees larger moment of inertia to resist this load. The cross-sectional area of the tree reduces gradually until it becomes effectively zero at the top. The sectional profile of the tree, again, stands in a better position to resist moment in the most efficient way. It is an economical design that has evolved over millions of years. It is the design that best suits the requirements of the surrounding (Load conditions).

Let us now consider another case when the tree is not subjected to wind or any lateral loads. The tree now acts as an isolated column. It is subjected to its own self-weight in vertically downward direction. This axial load increases from the top of the tree towards its bottom and so does the area of the cross-section. This helps the tree to resist the axial stress in the most optimized way. Smooth reduction of area of the cross-section also prevents stress concentration. Hence, this design is economical and structurally sound.

The concentration of mass in the stem ensures symmetry and stability of the structure. The branches of a tree can be observed growing symmetrically in all directions, providing for minimum eccentricity. It is often the case in many trees to have spread out branches for transpiration, breathing, photosynthesis and other biological purposes. The spread-out branches warrant higher surface area for the exposure to the sun, aiding in photosynthesis. It must be noted that the growth of these branches and the tree in general happens symmetrically reducing the eccentricity and maintaining the center of gravity of any section near the centerline of the tree.

The fibers running along the stem serves two purposes. The first being transportation of water and nutrition, and second being reinforcement. The continuous extension of the fibers from stem to branches guarantees fixed support (or very close to fixed support) at the joint. Any structure subjected to bending forces is subjected to axial stresses at the section. The fibers are aligned along the stem acting as reinforcements, to help bear the axial stresses. 
When a tree grows on an inclined terrain, it tries to straighten itself with its growth. This reduces the moment acting on the stem because of reduced lever arm distance. This evolving feature ensures that the center of gravity of the structure falls on the centerline of the stem where most of the mass is concentrated. Furthermore, it is observed that the roots of a tree grow in the direction opposite to the inclination of the tree. This provides proper anchorage to the tree by facilitating required fixed end moment at the bottom of tree (fixed support of cantilever).

\subsection{On material}
\begin{figure}[ht]\centering 
\includegraphics[width=10cm]{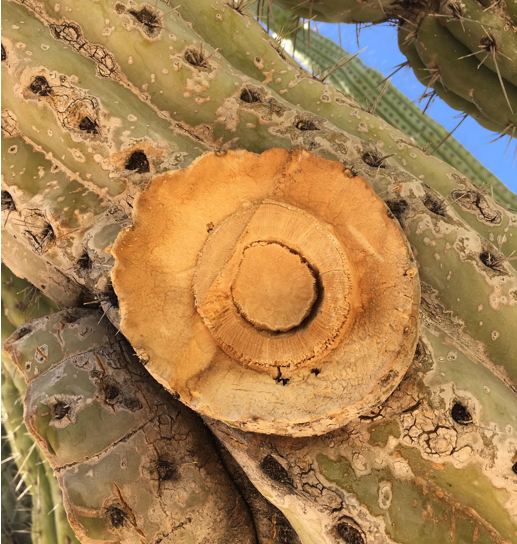}
\caption{Cross-section of the branch of cactus plant}
\label{fig:Figure2}
\end{figure}

Structural engineering is analogous to a triangle with material, geometry and load as its vertices. The tree is subjected to complex and extreme loads, some of which include its self-weight, snow load, wind load, and earthquake loads. Not only that, the requirements of tree differ with location, terrain, soil quality, sun light exposure, rainfall and many other factors. Structural engineers generally use materials with well-defined properties (desirably homogeneous and isotropic). It is difficult to maintain quality control in timber structures because of the inhomogeneous and anisotropic nature of wood. On the other hand, as far as quality control is concerned, steel is considered as one of the easiest material to operate with because of its homogeneous and isotropic nature. Unlike structural engineers, nature uses the best material irrespective of complexity in material properties. The wood has evolved with time being subjected to different conditions. Wood is a composite material that is highly anisotropic, inhomogeneous and organic. Since each tree is subjected to different surroundings, they grow differently. Each of them has a unique design agreeing to basic structural engineering requirements making each of them optimum and efficient. Figure \ref{fig:Figure2} shows the cross-section of cactus plant which is different from the cross-section of tree. Nature not only is just beautiful but also quite intelligent. 

\section{CONCLUSION}
This article details the structure of tree in conceptual way that highlights its optimized design. We, as structural engineers, know that there are large numbers of design solution for a given problem, but it can be realized that out of these solutions, there is just one most optimum and the most efficient solution. Nature certainly seems to follow that one design. This observation made us realize that nature is the best structural engineer.  

\end{document}